\documentclass[12pt]{article}
\usepackage{amsmath,epsfig,amsfonts}

\begin{document}
\hspace*{13cm}IU-MSTP/69 \\
\hspace*{13cm}hep-th/0506259 \\
\hspace*{13.9cm}June, 2005

\begin{center}
{\Large\bf Dirac Operator Zero-modes on a Torus}
\end{center}

\vspace*{1cm}
\def\thefootnote{\fnsymbol{footnote}}
\begin{center}{\sc Yasushi Tenjinbayashi${}^1$, Hiroshi Igarashi${}^2$ 
and Takanori Fujiwara${}^2$}
\end{center}
\vspace*{0.2cm}
\begin{center}
{\it ${}^1$ Graduate School of Science and Engineering, Ibaraki University, 
Mito 310-8512, Japan \\
${}^2$ Department of Physics, Ibaraki University, 
Mito 310-8512, Japan}
\end{center}

\vfill
\begin{center}
{\large\sc Abstract}
\end{center}
We study Dirac operator zero-modes on a torus for gauge background 
with uniform field strengths. Under the basic translations of the torus 
coordinates the wave functions are subject to twisted periodic conditions. 
In a suitable torus coordinates  the zero-mode wave functions can 
be related to holomorphic functions of the complex torus coordinates. 
We construct the zero-mode wave functions that satisfy the twisted 
periodic conditions. The chirality and the degeneracy of the zero-modes 
are uniquely determined by the gauge background and are consistent 
with the index theorem.

\noindent
\newpage
\pagestyle{plain}
\section{Introduction}
\label{sec:intro}
\setcounter{equation}{0}

Toric geometry appears in periodic systems. In many interesting cases 
background gauge field is also introduced. We may then consider 
topologically nontrivial gauge fields. They affects the spectrum of 
the hamiltonian in an interesting way even if the charged particle does 
not touch the magnetic field. In Dirac theory on a compact space we 
have chiral zero-modes for gauge background with nonvanishing 
topological charge. This is well-known as the Atiyah-Singer index 
theorem \cite{Atiyah-Singer}. It can be used as a mechanism of 
keeping fermions massless in Kaluza-Klein compactifications. 

If the gauge fields carry topological information of the periodic 
system, it cannot be periodic but changes by a gauge 
transformation under translations by a period of the system. 
Similarly the boundary condition for wave functions of a charged 
particle becomes twisted from the naive periodic boundary condition. 
This makes the task of solving eigenvalue problems nontrivial. 
In this paper we investigate Dirac fields in an abelian gauge 
background of constant field strength on a torus and construct 
zero-mode solutions of the Dirac operator. They are interesting in gauge 
theories in connection to chiral anomaly \cite{Adler,Nielsen-Ninomiya} and also in Kaluza-Klein 
scenario as a mechanism to have massless fermions. In Ref.  
\cite{Sakamoto-Tanimura} Sakamoto and Tanimura developed Fourier 
analysis for wave functions satisfying twisted boundary conditions 
on an arbitrary torus. They analyzed the eigenvalue problem of 
magnetic laplacian. Their arguments are based on the representation 
theory of magnetic translation group \cite{Brown, Zak,Tanimura}. 
Though their approach is powerful and systematic, it is possible 
to find Dirac operator zero-modes in a way that is more accessible 
to physicists. Namely, we directly solve massless Dirac equation 
on a euclidean torus and find solutions satisfying twisted boundary 
conditions. We show that  the zero-mode wave functions can 
be expressed as gaussian wave function times a function of 
complex torus coordinates. We find that the twisted 
boundary conditions determine the zero-mode wave 
functions up to an overall normalization. 

This paper is organized as follows. In the next section we 
argue general features of gauge potentials on a torus. We 
also show periodicity conditions for wave functions. In 
Sect. \ref{sec:mdeqot2} we solve the massless Dirac 
equation on $T^2$. This system exhibits some 
characteristics of Dirac operator zero-modes on tori. 
We show that the zero-mode wave functions are 
related to holomorphic or anti-holomprphic functions 
satisfying periodicity conditions with a twist. 
In Sect. \ref{sec:ethd} Dirac operator zero-modes in an 
arbitrary even dimensions are analyzed for field strengths 
of a standard form. Most general field strengths are treated 
in Sect. \ref{sec:etaedII}. The zero-mode wave functions 
can be related to holomorphic functions. We derive the 
periodicity conditions for the holomorphic functions. 
In Sect. \ref{sec:gpspc} we solve the periodicity 
conditions and give a construction of zero-mode wave 
functions. Sect. \ref{sec:sandd} is devoted to summary. 
A periodicity relation for general translations is 
derived in Appendix \ref{sec:gt}. In the appendices 
\ref{sec:fl} and \ref{sec:add2} we give proofs of the 
key mathematical facts that are used in this paper. 
The normalization constant of the zero-mode wave 
functions is given in Appendix \ref{sec:NIWF}.

\section{Gauge Potential on a Torus}
\label{sec:det2}
\setcounter{equation}{0}

A $d$ dimensional torus $T^d$ can be regarded 
as a quotient space ${\bf R}^d/\Lambda$, 
where $\Lambda$ is some $d$ dimensional 
lattice. Or equivalently, it can be obtained 
from a $d$ dimensional parallelogram by 
identifying its opposite $d-1$ dimensional 
faces. In this paper we take a hypercubic 
regular lattice $\Lambda={\bf Z}^d$ generated 
by the standard basis vectors $e_a$ 
($a=1,\cdots, d$) with $(e_a)^b=\delta_a^b$ and 
define $T^d={\bf R}^d/\Lambda^d$ by identifying 
all the points that differ by a vector in 
$\Lambda$. In Ref. \cite{Sakamoto-Tanimura} a more 
general torus is considered by taking an arbitrary set of 
vectors in place of the orthonormal basis $e_a$.  It 
is straightforward to carry out this kind of 
generalization in the following development. 

Abelian gauge fields on $T^d$ 
\begin{eqnarray}
  \label{eq:agf}
  A&=&\sum_aA_a(x){\rm d}x^a
\end{eqnarray}
are topologically classified by a set of 
integers $\phi_{ab}$ given by  
\begin{eqnarray}
  \label{eq:fluxes}
  \phi_{ab}&=&\frac{1}{2\pi}\int_{D_{ab}} F, 
\end{eqnarray}
where $D_{ab}$ is an $(a,b)$-plane taken in $T^d$ 
and $F$ is the field strength 2-form 
\begin{eqnarray}
  \label{eq:fs2f}
  F=\frac{1}{2}\sum_{a,b}F_{ab}
  {\rm d}x^a\wedge{\rm d}x^b={\rm d}A. 
\end{eqnarray}
We see that $2\pi\phi_{ab}$ is the flux through $D_{ab}$. 

\begin{figure}[htbp]
  \centering
  \epsfig{file=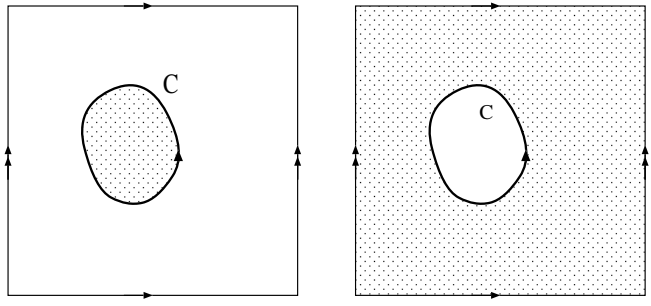,clip=,
    height=4cm,angle=0}
  \caption{Two possible choices of two-dimensional 
    surface with $C$ as its boundary. The square 
    denotes the $(a,b)$-plane $D_{ab}$ taken in 
    $T^d$.}
  \label{fig:paratr}
\end{figure}

That $\phi_{ab}$ must be an integer 
can be understood by considering a parallel 
transport of a wave function along a closed 
curve $C$ taken in $D_{ab}$ as depicted in 
Fig. \ref{fig:paratr}. The phase factor arising 
from the parallel transport along $C$ is given by 
\begin{eqnarray}
  \label{eq:phf}
  \exp\Biggl[-i\int_CA\Biggr].
\end{eqnarray}
In applying the Stokes theorem there are essentially 
two different possibilities of choosing a two-dimensional 
surface with $C$ as its boundary. One 
is the region enclosed by $C$, the shaded region 
of Fig. \ref{fig:paratr}(a). It is contractible to a point. 
The other is the shaded region of Fig. \ref{fig:paratr}(b). 
This is also an allowed surface since the left side of the 
square is identified with the right and the upper side with 
the lower. These two must yield the same phase factor. 
This implies 
\begin{eqnarray}
  \label{eq:cphf}
  \exp\Biggl[-i\int_{D_{ab}}F\Biggr]=\exp[
  -2\pi i\phi_{ab}]=1. 
\end{eqnarray}
We thus find that $\phi_{ab}$ must be an integer. 

The gauge potential $A$ cannot be single-valued 
on $T^d$ if it belongs to a topologically 
nontrivial sector. However, the field strength 
$F$ must be well-defined on $T^d$. In other 
words, it must satisfy the periodic boundary 
conditions
\begin{eqnarray}
  \label{eq:pbc}
  F_{ab}(x+e_c)&=&F_{ab}(x).
\end{eqnarray}
This implies that $A_a(x)$ and $A_a(x+e_b)$ 
can only differ by a gauge transformation 
as 
\begin{eqnarray}
  \label{eq:Atransl}
  A_a(x+e_b)&=&A_a(x)+\partial_a\lambda_b(x), 
\end{eqnarray}
where $\lambda_b$ may be a function on ${\bf R}^d$. 
This also gives rise to the transformation 
property of a wave function $\psi$ under 
the translations as
\begin{eqnarray}
  \label{eq:trprppsi}
  \psi(x+e_a)=g_a(x)\psi(x), \qquad
  g_a(x)=e^{i\lambda_a(x)},
\end{eqnarray}
where we are assuming that the covariant 
derivative is given by\footnote{We include 
the coupling constant in the definition 
of gauge potential.}  
\begin{eqnarray}
  \label{eq:covd}
  D_a\psi(x)&=&(\partial_a-iA_a(x))\psi(x).
\end{eqnarray}

\begin{figure}[htbp]
  \centering
  \epsfig{file=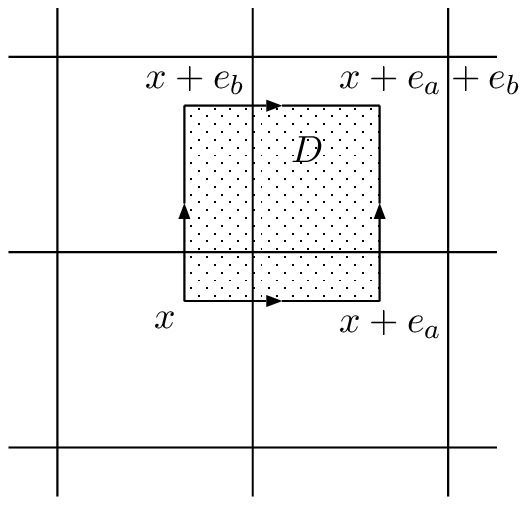,
    clip=,height=6cm,angle=0}
  \caption{}
  \label{fig:ttransl}
\end{figure}

In order for the translations 
(\ref{eq:trprppsi}) to give a unique wave 
function under the combined translations 
$x\rightarrow x+e_a \rightarrow x+e_a+e_b$ and 
$x\rightarrow x+e_b \rightarrow x+e_a+e_b$, the $g_a(x)$ 
must satisfy the cocycle conditions
\begin{eqnarray}
  \label{eq:cocyclec}
  g_a(x+e_b)g_b(x)=g_b(x+e_a)g_a(x).
\end{eqnarray}
That these are fulfilled can 
be seen by considering the parallel 
transport of the wave function along the 
boundary of the square $D$ as depicted in 
Fig. \ref{fig:ttransl}. 
The flux through $D$ can be computed as
\begin{eqnarray}
  \label{eq:fluxD}
  \int_D F=\int_{\partial D}A
  =\lambda_a(x+e_b)-\lambda_a(x)
  -\lambda_b(x+e_a)+\lambda_b(x).
\end{eqnarray}
In deriving this use has been made of 
(\ref{eq:Atransl}). Since the flux 
through $D$ is equal to $2\pi$ times an integer, 
we see that the cocycle conditions 
(\ref{eq:cocyclec}) are satisfied. 

We now turn to gauge potential with uniform field strength. 
In this case $F_{ab}$ can be written as  
\begin{eqnarray}
  \label{eq:Fabmbi}
  F_{ab}=2\pi \phi_{ab}. 
\end{eqnarray}
The gauge potential 1-form $A$ in axial gauge 
is given by 
\begin{eqnarray}
  \label{eq:gpiagc}
  A=\sum_{a<b}F_{ab}x^a{\rm d}x^b
  +\sum_b a_b{\rm d}x^b. 
\end{eqnarray}
Since we can freely shift the constant 
part $a_b$ by $2\pi$ by carrying out a gauge 
transformation on $T^d$ given by 
$g(x)=e^{2\pi i x^b}$, we can restrict $a_b$ 
to be in the interval $0\leq a_b<2\pi$. 
Furthermore, if $\det F_{ab}\ne 0$, we can 
remove the constant part of $A_a$ by a 
suitable shift of the coordinates $x$. This is not 
obvious in axial gauge. We come back to this 
point soon in connection to symmetric gauge. 
Henceforth, we restrict ourselves to 
the case $\det F_{ab}\ne0$ and choose the
coordinates so as for the gauge potential 
$A$ to be given by 
\begin{eqnarray}
  \label{eq:gpiag}
  A=\sum_{a<b}F_{ab}x^a{\rm d}x^b. 
\end{eqnarray}
This satisfies complete axial 
gauge conditions 
\begin{eqnarray}
  \label{eq:cagc}
  A_1(x_1,\cdots,x_d)=A_2(0,x_2,\cdots,x_d)
  =A_3(0,0,x_3,\cdots,x_d)=\cdots
  =A_d(0,\cdots,0,x_d)=0. \nonumber \\ 
\end{eqnarray}

The gauge potential (\ref{eq:gpiag}) 
satisfies the periodicity 
\begin{eqnarray}
  \label{eq:aperi}
  A(x+e_a)=A(x)+{\rm d}\lambda_a(x),
\end{eqnarray}
where $\lambda_a(x)$ is given by
\begin{eqnarray}
  \label{eq:lamacag}
  \lambda_a(x)=\sum_{b=a+1}^dF_{ab}x^b.
\end{eqnarray}
Due to (\ref{eq:Fabmbi}) the gauge transformation $g_a(x)$ in 
(\ref{eq:trprppsi}) is periodic in ${\bf R}^d$, {\it i.e. }, 
\begin{eqnarray}
  \label{eq:gperioa}
  g_a(x+e_b)=g_a(x).
\end{eqnarray}
We can regard it as a gauge transformation on $T^d$. 
The wave function $\psi$ must satisfy the periodicity
(\ref{eq:trprppsi}) with $\lambda_a$ given 
by (\ref{eq:lamacag}). 

Though the axial gauge is useful in our analysis, it lacks the 
rotational covariance. The shortcoming can be remedied by
using the gauge potential in symmetric gauge given by
\begin{eqnarray}
  \label{eq:gpisg}
  \tilde A_a(x)=\frac{1}{2}\sum_{b=1}^dF_{ba}x^b.  
\end{eqnarray}
It is related to the potential in axial gauge by a gauge transformation 
in ${\bf R}^d$ 
\begin{eqnarray}
  \label{eq:cag2sg}
  \tilde A_a=A_a+\partial_a\Lambda, \qquad
  \Lambda=-\frac{1}{2}\sum_{a<b}x^ax^bF_{ab}.
\end{eqnarray}
The wave function in symmetric gauge is related to the 
wave function in axial gauge by
\begin{eqnarray}
  \label{eq:cag2sgpsi}
  \tilde\psi(x)&=&e^{i\Lambda(x)}\psi(x).
\end{eqnarray}
The gauge potential (\ref{eq:gpisg}) is not of a most general 
form. We can add an arbitrary constant term to the potential
without affecting the field strength. It can be removed, however, 
by a suitable shift of the coordinates $x$ if the field strength 
satisfies $\det F_{ab}\ne0$.

\section{Massless Dirac Equation on $T^2$}
\label{sec:mdeqot2}
\setcounter{equation}{0}

To solve massless Dirac equation in a background gauge field with 
constant field strength we need several machinery. In two dimensions, 
however, the problem is rather simple but its solution is illuminating 
to understand higher dimensional cases.  In this section we give a 
construction of Dirac operator zero-modes on $T^2$ to illustrate 
some points that also apply in higher dimensional cases. 

The Dirac operator in two dimensions is given by 
\begin{eqnarray}
  \label{eq:doitd}
  D\hskip -.25cm/\hskip .1cm=\sum_{k=1,2}\sigma^k D_k
  =\left(
  \begin{matrix}
    0 & \partial_x-i\partial_y-i(A_x-iA_y) \\
    \partial_x+i\partial_y-i(A_x+iA_y) & 0
  \end{matrix}\right)
\end{eqnarray}
where $\sigma$'s are the Pauli matrices and the gauge potential 
for a constant field strength $F_{12}=B$ in 
axial gauge is given by 
\begin{eqnarray}
  \label{eq:ap2d}
  A_x=0, \qquad A_y=B x.
\end{eqnarray}
We can write $B=2\pi N\nu$, where $\nu$ is a positive integer
and $N={\rm sgn}B$. In two dimensions the distinction between 
$B$ and $2\pi \nu$ is not essential and the introduction of $\nu$ 
and $N$ might be considered redundant. As we will see in later 
sections, the distinction becomes crucial in the case of general 
uniform field strength in higher dimensions. 

The zero-mode of the Dirac operator is a solution to the 
euclidean massless Dirac equation 
\begin{eqnarray}
  \label{eq:deq}
  D\hskip -.25cm/\hskip .1cm\psi&=&0.
\end{eqnarray}
In terms of components 
\begin{eqnarray}
  \label{eq:comppsi}
  \psi&=&\left(
  \begin{matrix}
    \psi_+ \\ \psi_-
  \end{matrix}\right)
\end{eqnarray}
the Dirac equation (\ref{eq:deq}) is given by  
\begin{eqnarray}
  \label{eq:compdeq2}
  (\partial_x+is \partial_y+sB x)\psi_s=0. 
\end{eqnarray}
The $s=\pm$ label the chiral components of $\psi_s$ 
if we define the chiral matrix by $\Gamma_3=-i\sigma_1
\sigma_2=\sigma_3$. We see from (\ref{eq:compdeq2}) 
that $\psi_s$ can be written as 
\begin{eqnarray}
  \label{eq:ppm}
  \psi_s(x,y)&=&e^{-\frac{1}{2}sBx^2}f_s(x+isy), 
\end{eqnarray}
where $f_s$ is a function of $x+isy$ to be determined shortly. 

Under the translations $x\rightarrow x+e_a$ ($a=1,2$), 
the gauge potential (\ref{eq:ap2d}) is changed by 
a gauge transformation $\lambda_a(x)=By\delta_{a1}$. 
This gives the periodicity conditions on $\psi$ as 
\begin{eqnarray}
  \label{eq:pc2dp}
  \psi(x+1,y)=e^{iB y}\psi(x,y), \qquad 
  \psi(x,y+1)=\psi(x,y) 
\end{eqnarray}
These can be transcribed into the conditions for 
$f_s(x+isy)$ as 
\begin{eqnarray}
  \label{eq:fpix}
  && f_s(x+sN+isy)
  =e^{ 2\pi\nu\bigl(x+isy+\frac{1}{2}sN\bigr)}
  f_s(x+isy), \nonumber \\
  && f_s(x+is(y+N))=f_s(x+isy). 
\end{eqnarray}
Since $f_s(x+isy)$ are periodic under the translation 
$y\rightarrow y+N$, we can Fourier expand them as
\begin{eqnarray}
  \label{eq:Fexpf}
  f_s(x+isy)=\sum_nc^s_ne^{2\pi n(x+isy)}.
\end{eqnarray}
The Fourier coefficients $c^{s}_n$ can be determined from 
the first condition of (\ref{eq:fpix}). They must satisfy 
the recursion relation  
\begin{eqnarray}
  \label{eq:recrecc2d}
  c_n^s=e^{-\pi sN(2n-\nu)}c^s_{n-\nu}.  
\end{eqnarray}
By putting $n=\nu k+r$ with $k\in{\bf Z}$ and 
$r=0,1,\cdots,\nu-1$, these can 
be solved as
\begin{eqnarray}
  \label{eq:recsol2d}
  c_{\nu k+r}^s&=&c_r^s e^{-\pi sN(\nu k^2+2rk)}, 
\end{eqnarray}
where $c^s_r$ are arbitrary constants. 
The $f_s(x+isy)$ is now given by
\begin{eqnarray}
  \label{eq:fpmexpsol}
  f_s(x+isy)&=&\sum_{r=0}^{\nu-1}c_r^s
  \sum_{k=-\infty}^{+\infty}
  e^{-\pi sN(\nu k^2+2rk)+2\pi(\nu k+r)(x+isy)}.
\end{eqnarray}
Since $B=2\pi \nu N$, the sum defining $f_-$ does not converge 
if $B>0$. This implies 
that $c_r^-$ and, hence, $f_-$ must vanish. There are only $\nu$ 
independent $f_+$ corresponding to an arbitrary choice of 
$c_r^+$. We thus obtain $\nu$ independent zero-modes with 
positive chirality. Conversely, $f_+$ does 
not exists for $B<0$ and there are $\nu$ independent $f_-$, 
giving zero-modes with negative chirality. 

The absence of zero-mode of $D_1-iD_2$ for $B>0$ can be shown 
without solving the Dirac equation.  Using 
\begin{eqnarray}
  \label{eq:oprel}
  (D_1+iD_2)(D_1-iD_2)=D_1^2+D_2^2-B,
\end{eqnarray}
we get 
\begin{eqnarray}
  \label{eq:intzmsol}
  \int_{T^2} d^2x|(D_1-iD_2)\psi_-|^2
  =\int_{T^2} d^2x(|D_1\psi_-|^2+|D_2\psi_-|^2+B|\psi_-|^2).
\end{eqnarray}
The rhs of this expression cannot vanish if 
$\psi_-\ne0$. This implies that $D_1-iD_2$ has no 
nontrivial zero-modes for $B>0$. Conversely, $D_1+iD_2$ 
has no zero-mode if $B<0$.

Combining (\ref{eq:ppm}) and (\ref{eq:fpmexpsol}), we finally 
obtain the normalized zero-modes 
\begin{eqnarray}
  \label{eq:zms}
  &&\psi_{rR}=u_+\psi_{r,+}(x,y), \qquad(B>0) \nonumber \\
  &&\psi_{rL}=u_-\psi_{r,-}(x,y), \qquad(B<0)
\end{eqnarray}
where we have introduced the basis of two-component chiral 
spinors $u_\pm$ given by 
\begin{eqnarray}
  \label{eq:2csp}
  u_+=\left(
    \begin{matrix}
      1 \\0
    \end{matrix}\right), \qquad 
  u_-=\left(
    \begin{matrix}
      0 \\1
    \end{matrix}\right).
\end{eqnarray}
The $\psi_{r,s}(x,y)$ are defined by 
\begin{eqnarray}
  \label{eq:fs}
  \psi_{r,s}(x,y)&=&(2\nu)^{\frac{1}{4}}
  \sum_{k=-\infty}^{+\infty}e^{-\frac{\pi sN}{\nu}\bigl\{
    \nu(sNx-k)-r\Bigr\}^2+2\pi is(\nu k+r)y} \nonumber \\
  &=&(2\nu)^{\frac{1}{4}}
  \sum_{n\equiv r\:({\rm mod}\: \nu)}e^{-\frac{\pi sN}{\nu}(
    s\nu Nx-n)^2+2\pi isny},
\end{eqnarray}
where the last sum is taken over integers  of the form $n=\nu k+r$ 
($k\in{\bf Z}$). The overall factor is so chosen that the wave functions 
satisfy the orthonormality 
\begin{eqnarray}
  \label{eq:fson}
  \int_{T^2} d^2x\psi^\ast_{r,s}\psi_{r',s}=\delta_{r,r'}. 
\end{eqnarray}

The $\psi_{rR}$ and $\psi_{rL}$  are eigenstates of the chiral 
matrix $\Gamma_3$. If we denote the  number of positive and 
negative chirality zero-modes by $n_R$ and $n_L$ and define 
the index of $D\hskip -.25cm/\hskip .1cm$ 
by
\begin{eqnarray}
  \label{eq:asind}
  {\rm index}D\hskip -.25cm/\hskip .1cm=n_R-n_L, 
\end{eqnarray}
we find ${\rm index}D\hskip -.25cm/\hskip .1cm=\nu N=B/2\pi$ 
irrespective of the sign of $B$. This is completely consistent 
with the index theorem. 
\begin{eqnarray}
  \label{eq:2dindth}
  {\rm index}D\hskip -.25cm/\hskip .1cm
  =\frac{B}{2\pi}=\frac{1}{2\pi}\int_{T^2}F.
\end{eqnarray}

\section{Extension to Higher Dimensions}
\label{sec:ethd}
\setcounter{equation}{0}

The analysis given in the previous section can be easily extended 
to arbitrary even dimensions $d=2n$ in the case that the 
field strength 2-form  takes a special form 
\begin{eqnarray}
  \label{eq:sfs}
  F=\sum_{k=1}^nB_k {\rm d}x^{2k-1}\wedge{\rm d}x^{2k}, 
\end{eqnarray}
where $B_k$ is a flux density through $(2k-1,2k)$-plane 
and $\nu_k=|B_k|/2\pi$ is a positive integer. 
Most general uniform field strength can be made into the 
form (\ref{eq:sfs}) by a suitable coordinate rotation. 
In the rotated coordinates, however, the periodicity conditions 
(\ref{eq:trprppsi}) become complicated. We postpone the analysis 
of the Dirac operator zero-modes for the most general 
uniform field strength to later sections. In this section 
we argue the Dirac operator zero-modes for the special 
gauge background given by (\ref{eq:sfs}) and develop 
some machinery needed in later sections. 

The $\gamma$-matrices $\Gamma_n^a$ ($a=1,\cdots,d$) in 
$d=2n$ dimensions can be obtained recursively by the relations 
\begin{eqnarray}
  \label{eq:gmat}
  \Gamma_n^1=\sigma^1\otimes1_{2^{n-1}}, \quad
  \Gamma_n^2=\sigma^2\otimes1_{2^{n-1}}, \quad
  \Gamma_n^a=\sigma^3\otimes\Gamma_{n-1}^{a-2} , \qquad
  (a=3,\cdots,d),
\end{eqnarray}
where $1_N$ is the $N\times N$ unit matrix and $\Gamma_n^a$ are 
$2^n\times2^n$ matrices. Explicitly they are given by
\begin{eqnarray}
  \label{eq:gmat2}
  &&\Gamma^{2k+1}_n=\underbrace{\sigma^3\otimes\cdots\otimes
    \sigma^3\otimes}_{k~{\rm times}}
  \sigma^1\underbrace{\otimes1_2\otimes\cdots\otimes1_2}_{n-k-1~{\rm times}}, 
  \nonumber \\
  &&\Gamma^{2k+2}_n=\underbrace{\sigma^3\otimes\cdots\otimes
    \sigma^3\otimes}_{k~{\rm times}}
  \sigma^2\underbrace{\otimes1_2\otimes\cdots\otimes1_2}_{n-k-1~{\rm times}}.
\end{eqnarray}
The chiral matrix $\Gamma_{d+1}$ is given by 
\begin{eqnarray}
  \label{eq:cmat}
  \Gamma_{d+1}=(-i)^{n}\Gamma_n^1\cdots\Gamma_n^d
  &=&\underbrace{\sigma^3\otimes\cdots\otimes\sigma^3}_{n~{\rm times}}
\end{eqnarray}

The gauge potential that gives rise to the field strength 
(\ref{eq:sfs}) is a simple extension of (\ref{eq:ap2d}) and is given by 
\begin{eqnarray}
  \label{eq:gpotid}
  A_{2k-1}=0, \qquad A_{2k}=B_kx^{2k-1},
\end{eqnarray}
The Dirac operator is then given by
\begin{eqnarray}
  \label{eq:do}
  D\hskip -.25cm/\hskip .1cm=\Gamma^a_n(\partial_a-iA_a) 
  =\sum_{k=0}^{n-1}\sigma^3\otimes\cdots\otimes
    \sigma^3\otimes(\sigma^1D_{2k-1}+\sigma^2D_{2k})
    \otimes1_2\otimes\cdots\otimes1_2.
\end{eqnarray}
The components of the Dirac spinor can be defined by introducing  
a basis of the representation (\ref{eq:gmat2}) defined as 
direct products of two-component spinors $u_s$ ($s=\pm$) given by 
(\ref{eq:2csp}). They satisfy
\begin{eqnarray}
  \label{eq:brels}
  \sigma^3u_s=su_s, \qquad \sigma^+u_-=u_+, \qquad \sigma^-u_+=u_-
\end{eqnarray}
with $\sigma^\pm=\frac{1}{2}(\sigma^1\pm i\sigma^2)$. 
The basis of spinors in $d=2n$ dimensions is then given by  
\begin{eqnarray}
  \label{eq:us}
  u_{s_1}\otimes \cdots\otimes u_{s_n}.
\end{eqnarray}
Then an arbitrary spinor $\psi$ can be expanded  as 
\begin{eqnarray}
  \label{eq:dspexp}
  \psi(x)=\sum_{\{s\}}u_{s_1}\otimes \cdots\otimes u_{s_n}\psi_{s_1\cdots s_n}(x),
\end{eqnarray}
where $\psi_{s_1\cdots s_n}$ stand for the components of $\psi(x)$. 
Using $\sigma^1u_s=u_{-s}$ and $\sigma^2u_s=isu_{-s}$, 
we can write the Dirac equation in component form as 
\begin{eqnarray}
  \label{eq:deqicf}
  \sum_{k=1}^ns_1\cdots s_{k-1}(D_{2k-1}-is_kD_{2k})
  \psi_{s_1\cdots s_{k-1}-s_ks_{k+1}\cdots  s_n}(x)=0.
\end{eqnarray}
These are the generalization of (\ref{eq:compdeq2}) to $d=2n$ dimensions. 
Since each equation contains $n$ different components of $\psi(x)$, one might 
think it difficult to solve them. As we will see soon, this is not the case. 
We show that each component $\psi_{s_1\cdots s_n}$ must satisfy 
uncoupled first order differential equations similar to (\ref{eq:compdeq2}). 

To show this we first compute the square of the Dirac operator. By noting the 
commutation relations
\begin{eqnarray}
  \label{eq:Dcomm}
  [D_{2k-1},D_{2l-1}]=[D_{2k},D_{2l}]=0, \qquad
  [D_{2k-1},D_{2l}]=-iB_k\delta_{kl}, \qquad
  (k,l=1,\cdots,n)
\end{eqnarray}
we obtain 
\begin{eqnarray}
  \label{eq:sdop}
  D\hskip -.25cm/\hskip .1cm^2
  &=&D^2-i\sum_{k=1}^n\Gamma_n^{2k-1}\Gamma_n^{2k}B_k,
\end{eqnarray}
where we have introduced the Laplacian $D^2=\sum_aD_a^2$ . 
In the present representation of $\Gamma^a$ we have 
\begin{eqnarray}
  \label{eq:GG}
  \Gamma_n^{2k-1}\Gamma_n^{2k}
  =\underbrace{1_2\otimes\cdots\otimes1_2\otimes}_{k-1~{\rm times}}
  i\sigma^3\underbrace{\otimes1_2\otimes\cdots\otimes1_2}_{
    n-k~{\rm times}}.
\end{eqnarray}
This gives 
\begin{eqnarray}
  \label{eq:DsDs}
  D\hskip -.25cm/\hskip .1cm^2\psi&=&\sum_{\{s\}}u_{s_1}\otimes\cdots\otimes u_{s_n}
  \Bigl(D^2+\sum_{k=1}^ns_kB_k\Bigr)
  \psi_{s_1\cdots s_n}.
\end{eqnarray}
We thus find that each component of the Dirac operator zero-modes must satisfy 
\begin{eqnarray}
  \label{eq:D2}
  \Bigl(D^2+\sum_{k=1}^ns_kB_k\Bigr)
  \psi_{s_1\cdots s_n}&=&0.
\end{eqnarray}
By multiplying $\psi_{s_1\cdots s_n}^\ast$ to this expression 
and then integrating over $T^d$,  we obtain 
\begin{eqnarray}
  \label{eq:intDpDp}
  \int_{T^d} d^dx\sum_k|(D_{2k-1}+is_kD_{2k})\psi_{s_1\cdots s_n}|^2=0, 
\end{eqnarray}
where use has been made of the relation 
\begin{eqnarray}
  \label{eq:ufre}
  (D_{2k-1}-is_kD_{2k})(D_{2k-1}+is_kD_{2k})
  =D_{2k-1}^2+D_{2k}^2+s_kB_k.
\end{eqnarray}
We thus find that $\psi_{s_1\cdots s_n}$ must satisfy
\begin{eqnarray}
  \label{eq:DpiDp}
  (D_{2k-1}+is_kD_{2k})\psi_{s_1\cdots s_n}&=&0. 
\end{eqnarray}
These are essentially the same with the two-dimensional equations 
(\ref{eq:compdeq2}). As we have shown in the previous section, 
$D_{2k-1}+is_kD_{2k}$ has no zero-modes if ${\rm sgn} s_kB_k<0$. 
This implies that $\psi_{s_1\cdots s_n}=0$ unless $s_1={\rm sgn}B_1,~\cdots,~
s_n={\rm sgn}B_n$. The zero-modes of $D\hskip -.25cm/\hskip .1cm$ 
is then given by a tensor product as 
\begin{eqnarray}
  \label{eq:zm}
  \psi=\psi_1(x_1,x_2)\otimes\psi_2(x_3,x_4)\otimes\cdots\otimes\psi_n(x_{d-1},x_d),
\end{eqnarray}
where $\psi_k$ is a zero-mode of $\sigma^1D_{2k-1}+\sigma^2D_{2k}$. 
There are $\nu_k$ zero-modes of $\sigma^1D_{2k-1}+\sigma^2D_{2k}$ 
with the two-dimensional chirality ${\rm sgn}\,B_k$. 
The chirality of $\psi$ is the product of each chirality of the 
two-dimensional spinor $\psi_k$ and coincides with 
${\rm sgn}B_1\cdots B_n$. Since the number of independent 
zero-modes of $D\hskip -.25cm/\hskip .1cm$ is given by 
$\nu_1\nu_2\cdots \nu_n$, we obtain the index relation
\begin{eqnarray}
  \label{eq:indth}
  {\rm index}D\hskip -.25cm/\hskip .1cm
  =\prod_k\frac{B_k}{2\pi}=\frac{1}{(2\pi)^nn!}\int_{T^d} F^n.
\end{eqnarray}

\section{Extension to Arbitrary Field Strengths}
\label{sec:etaedII}
\setcounter{equation}{0}

In the previous section we have considered rather special gauge 
field that can be treated by the method developed for the two dimensional 
system described in Sect. \ref{sec:mdeqot2}. A general uniform field 
strength can be more complicated. The flux through $(i,j)$-plane may 
be nonvanishing for any $i\ne j$. Fortunately, an arbitrary uniform field 
strength $F_{ab}$ can be made into a standard form by an SO($d$) 
rotation as 
\begin{eqnarray}
  \label{eq:FeqRBR}
  F_{ab}=\sum_{c,d}\zeta_a^c\zeta_b^dB_{cd}, \qquad
  B=(B_{ab})=\left(
  \begin{matrix}
    0&B_1&&&& \\
    -B_1&0&&&& \\
    &&0&B_2&&& \\
    &&-B_2&0&&& \\
    &&&&\ddots&&&\\
    &&&&&0&B_n\\
    &&&&&-B_n&0
  \end{matrix}\right)  ,
\end{eqnarray}
where $R=(\zeta_a^b)$ is a real orthogonal matrix satisfying 
$RR^{\rm T}=R^{\rm T}R=1$ and $\det R=1$. 
Since we are assuming $\det F_{ab}\ne0$, any eigenvalue $B_k$ is 
nonvanishing. This ensures that the Dirac operator 
has a nonvanishing index. The essential difference between 
(\ref{eq:sfs}) and (\ref{eq:FeqRBR}) is that the eigenvalue $B_k$ in the 
latter is not necessarily a $2\pi$ multiple of an integer. We also note that 
the field strengths $F_{ab}$ can be expressed as
\begin{eqnarray}
  \label{eq:F2bzz}
  F_{ab}&=&\sum_kB_k(\zeta_a^{2k-1}\zeta_b^{2k}-\zeta_b^{2k-1}\zeta_a^{2k}).
\end{eqnarray}
We will frequently use the relation in the following development. 

The goal of this paper is to find the solutions to the Dirac equation coupled 
to the gauge potential in axial gauge (\ref{eq:gpiag}). It is given by 
\begin{eqnarray}
  \label{eq:Deqicagx}
  \sum_a\Gamma^a\Bigl(\partial_a+i\sum_{b<a}F_{ab}x^b\Bigr)\psi(x)&=&0.
\end{eqnarray}
What makes the task nontrivial is the periodicity conditions (\ref{eq:trprppsi}). 
In the present case they are given by 
\begin{eqnarray}
  \label{eq:psiagpc}
  \psi(x+e_a)&=&e^{i\sum_{b<a}F_{ab}x^b}\psi(x).
\end{eqnarray}
In Ref. \cite{Sakamoto-Tanimura} these are referred to as twisted periodicity 
conditions. We will see that they essentially determine the solution. 

In the axial gauge, however, the rotational symmetry is not manifest. So, we carry out 
the gauge transformation (\ref{eq:cag2sg}) and move in symmetric 
gauge. The Dirac equation (\ref{eq:Deqicagx}) then becomes
\begin{eqnarray}
  \label{eq:Deqisg}
  \sum_a\Gamma^a\Bigl(\partial_a+\frac{i}{2}\sum_bF_{ab}x^b\Bigr)
  \tilde\psi(x)&=&0. 
\end{eqnarray}
We now introduce new orthogonal coordinates $y^a$ by
\begin{eqnarray}
  \label{eq:ya}
  y^a=\sum_bx^b\zeta_b^a. 
\end{eqnarray}
The Dirac equation (\ref{eq:Deqisg}) can be converted to the form
\begin{eqnarray}
  \label{eq:PpDeq}
  \sum_a\Gamma^a\Bigl(\frac{\partial}{\partial y^a}
  +\frac{i}{2}\sum_bB_{ab}y^b\Bigr)\tilde\Psi(y)=0,
\end{eqnarray}
where $\tilde\Psi(y)=S\tilde\psi(x)$ with $S^\dagger\Gamma^aS=\sum_b
\Gamma^b\zeta_b^a$. The $S$ stands for the spinor representation 
of the SO($d$) rotation  $R$. 

We again go back to  axial 
gauge in the coordinate system $y$. This can be achieved by the 
inverse of the gauge transformation (\ref{eq:cag2sgpsi}). 
The Dirac equation (\ref{eq:Deqicagx}) finally becomes 
\begin{eqnarray}
  \label{eq:deicag}
  \sum_a\Gamma^a\Bigl(\frac{\partial}{\partial y^a}+i\sum_{b<a}B_{ab}y^b
  \Bigr)\Psi(y)=0, 
\end{eqnarray}
where $\Psi$ is related to the original $\psi$ by 
\begin{eqnarray}
  \label{eq:sgP2cagP}
  \Psi(y)=e^{\frac{i}{2}\sum_{a<b}y^ay^bB_{ab}-\frac{i}{2}\sum_{a<b}x^ax^bF_{ab}}
  S\psi(x).
\end{eqnarray}
The Dirac operator on the rhs of (\ref{eq:deicag}) is of the from 
given by (\ref{eq:do}). This implies that the Dirac equation can be 
reduced to the form 
\begin{eqnarray}
  \label{eq:zmeqfP}
  \Biggl(\frac{\partial}{\partial y^{2k-1}}+is_k\frac{\partial}{\partial y^{2k}}
  +s_kB_ky^{2k-1}\Biggr)\Psi_{s_1\cdots s_n}=0. \qquad 
  (k=1,\cdots,n)
\end{eqnarray}
We see that $\Psi_{s_1\cdots s_n}$ can be expressed as 
\begin{eqnarray}
  \label{eq:Psi2f}
  \Psi_{s_1\cdots s_n}(y)=e^{-\frac{1}{2}\sum_ks_kB_k(y^{2k-1})^2}
  f_{s_1\cdots s_n}(z),
\end{eqnarray}
where $f_{s_1\cdots s_n}$ depends only on the  $n$ complex variables 
$z^k=y^{2k-1}+is_ky^{2k}$ ($k=1,\cdots,n$).

We now turn to the periodicity conditions for the wave function
$\Psi$. The translation $x\rightarrow x+e_a$ corresponds to 
a translation of $y$ by a vector $\zeta_a=(\zeta_a^b)$ as 
can be seen from 
\begin{eqnarray}
  \label{eq:fundtr}
  y^b=\sum_cx^c\zeta_c^b\rightarrow 
  y^{\prime b}=\sum_c(x+e_a)^c\zeta_c^b=(y+\zeta_a)^b. 
\end{eqnarray}
Using the periodicity (\ref{eq:psiagpc}) under the translation $x
\rightarrow x+e_a$, we can find the periodicity of $\Psi$ under the translation 
(\ref{eq:fundtr}) as 
\begin{eqnarray}
  \label{eq:Psitr}
  \Psi(y+\zeta_a)&=&e^{i\sum_{b<c}\zeta_a^bB_{bc}\bigl(y^c
  +\frac{1}{2}\zeta_a^c\bigr)}\Psi(y).
\end{eqnarray}
The overall factor of gauge transformation 
\begin{eqnarray}
  \label{eq:oafgt}
  g_a(y)&=&e^{i\sum_{b<c}\zeta_a^bB_{bc}\bigl(y^c
  +\frac{1}{2}\zeta_a^c\bigr)}
\end{eqnarray}
satisfies the cocycle conditions analogous to (\ref{eq:cocyclec}).  
This can be shown explicitly by noting the relation (\ref{eq:F2bzz}). 
The $g_a(y)$, however, is not single-valued on $T^d$. This is contrasted 
with the corresponding factor appearing in (\ref{eq:psiagpc}), where 
the gauge transformation is periodic under $x\rightarrow x+e_b$. 

Under the translation (\ref{eq:fundtr}) the $z^k$ are transformed as 
\begin{eqnarray}
  \label{eq:translz}
  z^k\rightarrow z^k+\Xi^k_a \qquad 
  \hbox{with}\qquad 
  \Xi_a^k=\zeta_a^{2k-1}+is_k\zeta_a^{2k}.
\end{eqnarray}
The periodicity conditions (\ref{eq:Psitr}) can be transcribed as the 
conditions for $f_{z_1\cdots s_n}(z)$. They are given by 
\begin{eqnarray}
  \label{eq:pc}
  f_{s_1\cdots s_n}(z+\Xi_a)=e^{\sum_ks_kB_k\zeta_a^{2k-1}(z^k
    +\frac{1}{2}\Xi_a^k)}f_{s_1\cdots s_n}(z).
\end{eqnarray}
These ensure that the wave functions defined in 
${\bf R}^d$ can be regarded as those on $T^d$. We will see that they 
imposes very tight constraints on the wave functions and are 
sufficient to determine $f_{s_1\cdots s_n}$. 

Before closing this section we generalize (\ref{eq:pc})  to an arbitrary 
translation by a vector $\Xi_a=\sum_aN^a\Xi_a$, where $N^a$ 
($a=1,\cdots,2n$) are integers. This can be achieved 
by applying (\ref{eq:pc}) repeatedly to $f_{s_1\cdots s_n}(z+\Xi)$ until 
we arrive at $f_{s_1\cdots s_n}(z)$. Since we will use a similar formula 
in later sections we give the periodicity conditions for an arbitrary translation 
in Appendix \ref{sec:gt}. They are given by 
\begin{eqnarray}
  \label{eq:fXitr}
  f_{s_1\cdots s_n}(z+\Xi)=e^{\sum_ks_k\zeta^{2k-1}B_k\bigl(z^k+\frac{1}{2}\Xi^k\bigr)
    -\frac{i}{2}\sum_{a<b}N^aN^bF_{ab}}f_{s_1\cdots s_n}(z), 
\end{eqnarray}
where $\zeta^{2k-1}=\sum_aN^a\zeta_a^{2k-1}$ is the real part of $\Xi_a^k$. 

The periodicity under the translation 
$z\rightarrow z+\Xi$ becomes much simpler if we can find a set of integers 
$N^a$ for which $\zeta^{2k-1}$ ($k=1,\cdots,n$) vanish. In the next section we 
show that we can find a set of translations under which $f_{s_1\cdots s_n}(z)$ is 
periodic or anti-periodic.

\section{Solving the Periodicity Conditions}
\label{sec:gpspc}
\setcounter{equation}{0}

In this section we give a construction of $f_{s_1\cdots s_n}(z)$. To 
achieve this we search for a basis vectors other than $\{e_a\}$ for the 
lattice ${\bf Z}^{2n}$ and convert the periodicity conditions 
(\ref{eq:pc}) into a tractable form. 

We first note that the field strengths $F_{ab}$ 
can be made into a block-diagonal form 
\begin{eqnarray}
  \label{eq:FLLNU}
  \nu_{ab}=\frac{1}{2\pi} L_a{}^cL_b{}^dF_{ab}
  =\left(
  \begin{matrix}
    0&\nu_1&&&& \\
    -\nu_1&0&&&& \\
    &&0&\nu_2&&& \\
    &&-\nu_2&0&&& \\
    &&&&\ddots&&&\\
    &&&&&0&\nu_n\\
    &&&&&-\nu_n&0
  \end{matrix}\right), 
\end{eqnarray}
where $L=(L_a{}^b)$ is a regular $2n\times2n$ matrix with integral 
elements satisfying $|\det L|=1$ and $\nu_p$ ($p=1,\cdots,n$) are 
positive integers with the property that $\nu_p$ divides 
$\nu_q$ for $p<q$. As before, we are assuming a general field 
strengths $F_{ab}$ with $\det F_{ab}\ne 0$ and any $\nu_p$ is 
nonvanishing. For given $F_{ab}$ we can find $L$ in a systematic way. 
To illustrate this we give a proof of this lemma \cite{Igusa} in 
Appendix \ref{sec:fl}.

We denote $L$ by $2n$ vectors $M_p=(M_p{}^a)$ 
and $N_p=(N_p{}^a)$ as
\begin{eqnarray}
  \label{eq:L}
  L=(M_1,N_1,\cdots,M_n,N_n). 
\end{eqnarray}
Then the relation (\ref{eq:FLLNU}) can be expressed as
\begin{eqnarray}
  \label{eq:FMN}
  F(M_p,N_q)=\nu_p\delta_{pq}, \qquad 
  F(M_p,M_q)=F(N_p,N_q)=0,  
\end{eqnarray}
where we have introduced a skew symmetric bilinear form 
\begin{eqnarray}
  \label{eq:Fxieta}
  F(\xi,\eta)=\frac{1}{2\pi}\sum_{a,b}F_{ab}\xi^a\eta^b. 
\end{eqnarray}
It is integer-valued if both $\xi$ and $\eta$ are integral vectors.

As we show in Appendix \ref{sec:fl}, any vector in 
$\Lambda={\bf Z}^d$ can be uniquely expressed by a 
linear combination of the $2n$ vectors $M_p$ and $N_p$ 
with integral coefficients. This implies that $\Lambda$ can also be 
generated by the basis $\{M_p,N_p\}$. In defining 
$T^{2n}={\bf R}^{2n}/{\bf Z}^{2n}$, 
the $2n$ dimensional hypercubic region $0\leq x^a\leq 1$ in ${\bf R}^{2n}$ 
can be replaced with the rectangular region $D$ given by
\begin{eqnarray}
  \label{eq:x2xieta}
  x^a=\sum_p(M_p{}^a\xi^p+N_p{}^a\eta^p).  \qquad
  (0\leq \xi^p,~\eta^p\leq1)
\end{eqnarray}
Using (\ref{eq:FMN}), we can solve this with respect to $\xi^p$ and $\eta^p$ as
\begin{eqnarray}
  \label{eq:xieta2x}
  \xi^p=\frac{F(x,N_p)}{\nu_p}, \qquad
  \eta^p=\frac{F(M_p,x)}{\nu_p}.
\end{eqnarray}
The volume of $D$ can be easily found as
\begin{eqnarray}
  \label{eq:vD}
  \int_Dd^{2n}x=\int_0^1d^n\xi\int_0^1d^n\eta|\det(M_p{}^a,N_p{}^a)|=1, 
\end{eqnarray}
where use has been made of $|\det L|=1$. These imply that $\xi^p$ and $\eta^p$ 
can be regarded as the coordinates of the $2n$-torus $T^{2n}$. Later in this section 
we consider the translations 
\begin{eqnarray}
  \label{eq:xtrMN}
  x^a\rightarrow x^a+M_p{}^a,~~N_p{}^a. 
\end{eqnarray}
These correspond to unit translations of the $(\xi,\eta)$-coordinates.  

In Sect. \ref{sec:etaedII}, we have used the fact that the field 
strength can be made into a standard form (\ref{eq:FeqRBR}) 
by an SO($2n$) ration $R=(\zeta_a^b)$. The choice of $R$, 
however, is not unique but we can replace $\zeta_a^{2k-1}$ 
and $\zeta_a{}^{2k}$ by 
\begin{eqnarray}
  \label{eq:zetap}
  \zeta_a^{2k-1}\cos\theta_k+\zeta_a^{2k}\sin\theta_k, 
  \qquad -\zeta_a^{2k-1}\sin\theta_k+\zeta_a^{2k}\cos\theta_k, 
\end{eqnarray}
where $\theta_k$ is an arbitrary rotation angle. By utilizing 
this arbitrariness we can always choose $\zeta_a^b$ to satisfy
\begin{eqnarray}
  \label{eq:nzetaeq0}
  \sum_aN_p{}^a\zeta_a^{2k-1}&=&0, \qquad
  (k,p=1,\cdots,n)
\end{eqnarray}
where $N_p=(N_p{}^a)$ are those vectors that constitute 
the $L$ as given by (\ref{eq:L}). In Appendix \ref{sec:add2}, 
we give a proof of the existence of such a basis 
$\zeta^b=(\zeta^b_a)$.

For the choice of $\zeta_a^b$ satisfying (\ref{eq:nzetaeq0}) 
the periodicity (\ref{eq:pc}) under the translations 
\begin{eqnarray}
  \label{eq:Omega}
  z\quad \rightarrow \quad z+\Omega_p \qquad
  \hbox{with}\qquad \Omega_p^k=is_k\sum_aN_p{}^a\zeta_a^{2k}
\end{eqnarray}
reduces to 
\begin{eqnarray}
  \label{eq:omgp}
  f_{s_1\cdots s_n}(z+\Omega_p)=e^{-i\pi\epsilon_p}f_{s_1\cdots s_n}(z), 
\end{eqnarray}
where $\epsilon_p$ is defined by
\begin{eqnarray}
  \label{eq:ep}
  \epsilon_p=\frac{1}{2\pi}\sum_{a<b}N_p{}^aN_p{}^bF_{ab}.
\end{eqnarray}
There are exactly $n$ such translations corresponding 
$N_p$ ($p=1,\cdots,n$). Since $\epsilon_p$ is an integer, we see that 
$f_{s_1\cdots s_n}(z)$ is either periodic or anti-periodic under the 
translations (\ref{eq:Omega}). 

It is convenient to introduce new variables $w^p$ by 
\begin{eqnarray}
  \label{eq:z2w}
  z^k=\sum_p\sum_as_kN_p{}^a\zeta_a^{2k}w^p. 
\end{eqnarray}
By noting (\ref{eq:F2bzz}) and (\ref{eq:nzetaeq0}), this can be 
made into the form 
\begin{eqnarray}
  \label{eq:z2w-2}
  \sum_ks_kB_k\zeta_a^{2k-1}z^k&=&2\pi\sum_ps_{ap}w^p, 
\end{eqnarray}
where $s_{ap}$ is an integer defined by
\begin{eqnarray}
  \label{eq:sap}
  s_{ap}=\frac{1}{2\pi}\sum_bF_{ab}N_p{}^b.
\end{eqnarray}
From (\ref{eq:FMN}) we see that $s_{ap}$ satisfies the orthogonality 
relation 
\begin{eqnarray}
  \label{eq:Ms}
  \sum_aM_p{}^as_{qa}=\nu_p\delta_{pq}.
\end{eqnarray}
This enables us  to solve (\ref{eq:z2w-2}) with respect to $w^p$ as 
\begin{eqnarray}
  \label{eq:wp}
  w^p&=&\frac{1}{2\pi\nu_p}\sum_k\sum_as_kB_kM_p{}^a\zeta_a^{2k-1}z^k.
\end{eqnarray}
The translation $z\rightarrow z+\Omega_p$ corresponds to a unit 
imaginary translation in $w$
\begin{eqnarray}
  \label{eq:trpwp}
  w\rightarrow w+ie_p,
\end{eqnarray}
where $e_p=(\delta_p^q)$ is a unit vector. 
This can be seen easily as 
\begin{eqnarray}
  \label{eq:ztr2}
  z^k\quad\rightarrow\quad  z'{}^k=\sum_q\sum_as_kN_q{}^a\zeta_a^{2k}
  (w^q+i\delta^q_p) =z^k+\Omega_p^k.
\end{eqnarray}
If we denote $f_{s_1\cdots s_n}(z)$ by $g_{s_1\cdots s_n}(w)$, the periodicity (\ref{eq:omgp}) can be
written as  
\begin{eqnarray}
  \label{eq:pag}
  g_{s_1\cdots s_n}(w+ie_p)=e^{-i\pi\epsilon_p}g_{s_1\cdots s_n}(w).
\end{eqnarray}

Since the change of variables (\ref{eq:z2w}) is nonsingular, 
the shift of $w\rightarrow w+\Upsilon_a$ under the translation 
$z\rightarrow z+\Xi_a$ is uniquely determined as 
\begin{eqnarray}
  \label{eq:upsa}
  \Upsilon_a^p=\frac{1}{2\pi\nu_p}\sum_b\sum_ks_kB_kM_p{}^b\zeta_b^{2k-1}
  \Xi_a^k.
\end{eqnarray}
We thus obtain from (\ref{eq:pc}) the periodicity relations of $g_{s_1\cdots s_n}(w)$ 
under $w\rightarrow w+\Upsilon_a$ as 
\begin{eqnarray}
  \label{eq:fpcg}
  g_{s_1\cdots s_n}(w+\Upsilon_a)=e^{2\pi\sum_{p}s_{ap}\bigl(w^p
    +\frac{1}{2}\Upsilon_a^p\bigr)}g_{s_1\cdots s_n}(w), 
\end{eqnarray}
where use has been made of (\ref{eq:z2w-2}). 

We can generalize (\ref{eq:fpcg}) for arbitrary $\Upsilon=\sum_aM^a\Upsilon_a$ 
as in (\ref{eq:fXitr}). It is given by 
\begin{eqnarray}
  \label{eq:genMg}
  g_{s_1\cdots s_n}(w+\Upsilon)=e^{2\pi\sum_p\sum_aM^as_{ap}\bigl(w^p+\frac{1}{2}
    \Upsilon^p\bigr)
    +\frac{i}{2}\sum_{a<b}M^aM^bF_{ab}}g_{s_1\cdots s_n}(w), 
\end{eqnarray}
where $M^a$ are arbitrary integers. In particular under the translation 
\begin{eqnarray}
  \label{eq:rho}
  w\rightarrow w+\rho_p \qquad \hbox{with}\qquad 
  \rho_p{}^q=\sum_a M_p{}^a\Upsilon_a^q, 
\end{eqnarray}
where $M_p=(M_p{}^a)$ is a vector  appearing in $L$, the wave function 
is transformed as 
\begin{eqnarray}
  \label{eq:gwu}
  g_{s_1\cdots s_n}(w+\rho_p)&=&e^{2\pi\nu_p\bigl(w^p+\frac{1}{2}\rho_p{}^p\bigr)
    +i\pi\tilde\epsilon_p\nu_p}g_{s_1\cdots s_n}(w),
\end{eqnarray}
In deriving this use has been made of (\ref{eq:Ms}). The $\tilde\epsilon_p$ is defined by  
\begin{eqnarray}
  \label{eq:et}
  \tilde\epsilon_p
  =\frac{1}{2\pi\nu_p}\sum_{a<b}M_p{}^aM_p{}^bF_{ab}. 
\end{eqnarray}
Note that only $w_p$ appears in the exponent on the rhs of (\ref{eq:gwu}) 
under the translation (\ref{eq:rho}). 

We are now in a position to solve the periodicity conditions (\ref{eq:pag}) and (\ref{eq:gwu}). 
From the periodicity (\ref{eq:pag}) we can express $g_{s_1\cdots s_n}(w)$ in the Fourier 
expansion as
\begin{eqnarray}
  \label{eq:gfexp}
  g_{s_1\cdots s_n}(w)=\sum_{\{n_p\}}c_{\{n_p\}}^{s_1\cdots s_n}
  e^{2\pi\sum_p(n_p-\frac{1}{2}\epsilon_p)w^p}, 
\end{eqnarray} 
where $\{n_p\}$ stands for the abbreviation of $n_1\cdots n_n$. 

The condition (\ref{eq:gwu}) can be translated into the recursion relations for the 
Fourier coefficients. They are given by 
\begin{eqnarray}
  \label{eq:rrrelc}
  c_{\{n_q\}}^{s_1\cdots s_n}
  =c_{\{n_q-\nu_p\delta_{pq}\}}^{s_1\cdots s_n}e^{-\pi\sum_q\rho_p{}^q(2n_q-\epsilon_q)
    +\pi\nu_p\rho_p{}^p+i\pi\tilde\epsilon_p\nu_p}.
\end{eqnarray}

It is straightforward to obtain the general expression for $c_{\{n_q\}}^{s_1\cdots s_n}$. 
We find that $c_{\{n_q\}}^{s_1\cdots s_n}$ for $n_q=\nu_qk_q+r_q$ 
with $k_q\in{\bf Z}$ and $r_q=0,1,\cdots,\nu_q-1$ is given by 
\begin{eqnarray}
  \label{eq:cn}
  c_{n_1\cdots n_n}^{s_1\cdots s_n}&=&c_{r_1\cdots r_n}^{s_1\cdots s_n} 
  e^{-\pi\sum_{p,q}\rho_p{}^q\nu_qk_pk_q
    -\pi\sum_{p,q}\rho_p{}^qk_p(2r_q-\epsilon_q)
    +i\pi\sum_p\tilde\epsilon_p\nu_pk_p}, 
\end{eqnarray}
where $c_{r_1\cdots r_n}^{s_1\cdots s_n}$ are arbitrary constants to be determined later. 
In deriving this we have used the relation
\begin{eqnarray}
  \label{eq:symprho}
  \rho_p{}^q\nu_q-\rho_q{}^p\nu_p=-iF(M_p,M_q)=0.
\end{eqnarray}

To each component of the wave function labeled by  $\{s_k\}$ we have 
$\nu_1\cdots\nu_n$ independent solutions to the recursion relations (\ref{eq:cn}).
They do not, however, necessarily correspond to nontrivial zero-mode solutions. 
The Fourier expansion (\ref{eq:gfexp}) must converge. 
To establish the convergence of the rhs of (\ref{eq:gfexp}), we 
must show that $c_{n_1\cdots n_n}^{s_1\cdots s_n}
\rightarrow0$ as $|n|\rightarrow\infty$, where 
$|n|=\sqrt{n_1^2+\cdots+n_n^2}$. It suffices to verify the convergence 
of the most dominant part 
\begin{eqnarray}
  \label{eq:mdp}
  |e^{-\pi\sum_{p,q}\rho_p{}^q\nu_qk_pk_q}|^2
  &=&e^{-\sum_ks_kB_k\bigl(\sum_{p}\sum_a\zeta_a^{2k-1}M_p{}^ak_p\bigr)^2}.
\end{eqnarray}
Since we have $\det\sum_a\zeta_a^{2k-1}M_p{}^a\ne0$, the rhs does not converge as 
$|n|\rightarrow\infty$ unless $s_kB_k>0$ for $k=1,\cdots,n$. This implies that the 
coefficients $c_{r_1\cdots r_n}^{s_1\cdots s_n}$ must vanish if $\{s_1,\cdots,s_n\}\ne
\{{\rm sgn}B_1,\cdots,{\rm sgn}B_n\}$. We thus find that the only nontrivial components of 
the zero-modes is $\Psi_{s_1\cdots s_n}$ with $s_k={\rm sgn}(B_k)$. 
Hence the number of independent zero-modes is equal to the number of 
arbitrary constants $c^{s_1\cdots s_n}_{r_1\cdots r_n}$ and is given by 
\begin{eqnarray}
  \label{eq:prodnu}
  \nu_1\cdots\nu_n
  =\sqrt{\det \Biggl(\frac{1}{2\pi}F_{ab}\Biggr)}
  =\frac{1}{(2\pi)^nn!}|\epsilon_{a_1b_1\cdots a_nb_n}F_{a_1b_1}\cdots F_{a_nb_n}|, 
\end{eqnarray}
where $\epsilon_{a_1b_1\cdots a_nb_n}$ is the Levi-Civita symbol in $d=2n$ dimensions. 
Furthermore, the chirality of these zero-modes coincides with 
$s_1\cdots s_n$. These are perfectly consistent with the index 
theorem (\ref{eq:indth}). 

We now turn to the wave function (\ref{eq:Psi2f}). To write it 
in terms of the present coordinates let us denote $w^p$ by its 
real and imaginary parts as 
\begin{eqnarray}
  \label{eq:uv}
  w^p&=&u^p+iv^p. \qquad (p=1,\cdots,n)
\end{eqnarray}
Then we have from (\ref{eq:wp})
\begin{eqnarray}
  \label{eq:guu}
  \sum_ks_kB_k(y^{2k-1})^2&=&2\pi\sum_{p,q}g_{pq}u^pu^q,
\end{eqnarray}
where $g_{pq}$ is given by 
\begin{eqnarray}
  \label{eq:gpq}
  g_{pq}&=&\frac{1}{2\pi}\sum_{a,b}\sum_ks_kB_k\zeta_a^{2k}\zeta_b^{2k}
  N_p{}^aN_q{}^b.
\end{eqnarray}
It reminds us of a metric in geometry. 
To find the inverse of $g_{pq}$ we note the relations
\begin{eqnarray}
  \label{eq:sNsB}
  && \sum_k\Bigl(\sum_as_kN_p{}^a\zeta_a^{2k}\Bigr)
  \Bigl(\sum_bs_kB_kM_q{}^b\zeta_b^{2k-1}\Bigr)=
  2\pi\nu_p\delta_{pq}, \\
  \label{eq:sBsN}
  && \sum_p\Bigl(\frac{1}{2\pi \nu_p}\sum_as_kB_kM_p{}^a\zeta_a^{2k-1}\Bigr)
  \Bigl(\sum_bs_lN_p{}^b\zeta_b^{2l}\Bigr)=\delta^{kl}.
\end{eqnarray}
These can be verified by using (\ref{eq:F2bzz}), (\ref{eq:FMN}) and (\ref{eq:nzetaeq0}). 
The inverse $g^{pq}$ can be obtained from (\ref{eq:gpq}) and 
(\ref{eq:sBsN}) as 
\begin{eqnarray}
  \label{eq:im}
  g^{pq}&=&\sum_{a,b}\sum_k\frac{s_kB_k}{2\pi\nu_p\nu_q}
  \zeta_a^{2k-1}\zeta_b^{2k-1}M_p{}^aM_q{}^b.
\end{eqnarray}
We see from (\ref{eq:upsa}) and (\ref{eq:rho}) that $g^{pq}$ is the 
real part of $\rho_p{}^q/\nu_p$. We write $\rho_p{}^q$ as 
\begin{eqnarray}
  \label{eq:rho2}
  \rho_p{}^q&=&\nu_p(g^{pq}+i\gamma^{pq}),
\end{eqnarray}
where $\gamma^{pq}$ is defined by 
\begin{eqnarray}
  \label{eq:gamma}
    \gamma^{pq}=\frac{1}{4\pi\nu_p\nu_q}\sum_{a,b}
  \sum_kB_k(\zeta_a^{2k-1}\zeta_b^{2k}+\zeta_b^{2k-1}\zeta_a^{2k})M_p{}^aM_q{}^b. 
\end{eqnarray}
The wave function (\ref{eq:Psi2f}) for a given set of $\{r_p\}$ can be written as
\begin{eqnarray}
  \label{eq:Psis}
  \Psi^{r_1\cdots r_n}_{s_1\cdots s_n}&=&
  \sum_{\overline n_1,\cdots,\overline n_n}
  e^{-i\pi\sum_{p,q}\gamma^{pq} 
  \overline n_p\overline n_q
  +i\pi\sum_p\tilde\epsilon_p\overline n_p
  -\pi\sum_{p,q}g_{pq}\bigl(u^p+\sum_rg^{pr}\overline n_r\bigr)
  \bigl(u^q+\sum_sg^{qs}\overline n_s\bigr)
  +2\pi i\sum_p\overline n_pv^p}, \nonumber \\ 
\end{eqnarray}
where the sum with respect to $\overline n_p$ ($p=1,\cdots,n$) should be taken over 
the numbers of the form $\overline n_p=\nu_pk_p+r_p-\frac{1}{2}\epsilon_p$ 
($k_p=0,\pm1,\pm2,\cdots$). 
We have suppressed overall normalization constant in (\ref{eq:Psis}).

The wave function (\ref{eq:Psis}) is not the final result. We must go back to 
the wave function $\psi(x)$ in the original coordinates. As we have mentioned 
at the beginning of this section, $\xi^p$ and $\eta^p$ defined by (\ref{eq:x2xieta}) 
are natural coordinates on $T^{2n}$. The coordinates $u^p$ and $v^p$ are 
related to  $\xi^p$ and $\eta^p$ by 
\begin{eqnarray}
  \label{eq:xie2uv}
  u^p=\sum_qg^{pq}\nu_q\xi^q, \qquad v^p=\eta^p+\sum_q\gamma^{pq}\nu_q\xi^q. 
\end{eqnarray}
The translations (\ref{eq:trpwp}) and (\ref{eq:rho}) correspond to
\begin{eqnarray}
  \label{eq:tpxe}
  \xi^q\rightarrow\xi^q, \quad \eta^q\rightarrow\eta^q+\delta^q_p \quad
  \hbox{under}\quad w^q\rightarrow w^q+i\delta^q_p, \nonumber \\
  \xi^q\rightarrow\xi^q+\delta^q_p, \quad \eta^q\rightarrow\eta^q \quad
  \hbox{under}\quad w^q\rightarrow w^q+\rho_p{}^q.
\end{eqnarray}
These can be understood by recalling the fact that the translation 
(\ref{eq:trpwp}) and (\ref{eq:rho}), respectively, correspond to 
$x^a\rightarrow x^a+N_p{}^a$ and $x^a\rightarrow x^a+M_p{}^a$. 
We thus find that  (\ref{eq:trpwp}) and (\ref{eq:rho}) are the unit 
translations of the coordinates $\xi^p$ and $\eta^p$. These are just the 
basic moves on $T^{2n}$. We can also directly show (\ref{eq:tpxe}) by 
making use of (\ref{eq:xie2uv}).

It is now possible to write down the zero-mode wave function $\psi(x)$ in terms 
of $\xi^p$ and $\eta^p$. Since the spinor part of $\Psi$ is given by 
$u_{s_1}\otimes\cdots\otimes u_{s_n}$, the normalized zero-mode for a given 
$\{r_p\}$ can be obtained from (\ref{eq:sgP2cagP}) and (\ref{eq:Psis}) as 
\begin{eqnarray}
  \label{eq:zmwfix}
  \psi^{r_1\cdots r_n}(x)
  &=&
  \Biggl[\frac{2^{\frac{n}{2}}\nu_1\cdots\nu_n}{\sqrt{\det g_{pq}}}\Biggr]^{\frac{1}{2}}
  S^\dagger u_{s_1}\otimes\cdots\otimes u_{s_n} F_{r_1\cdots r_n}(x)
  e^{\frac{i}{2}\sum_{a<b}F_{ab}x^ax^b}, 
\end{eqnarray}
where $F_{s_1\cdots s_n}^{r_1\cdots r_n}(x)$ is given by 
\begin{eqnarray}
  \label{eq:F}
   F_{s_1\cdots s_n}^{r_1\cdots r_n}(x)&=&e^{-\frac{i}{2}\sum_{a<b}B_{ab}y^ay^b}
   \Psi^{r_1\cdots r_n}_{s_1\cdots s_n}(y) \nonumber \\
   &=&\sum_{\overline n_1,\cdots \overline n_n}
   e^{-\pi\sum_{p,q}(g^{pq}+i\gamma^{pq})(\nu_p\xi^p-\overline n_p)(\nu_q\xi^q-\overline n_q)
  -i\pi\sum_p(\nu_p\xi^p-\overline n_p)\eta^p
  +i\pi\sum_p\overline n_p\eta^p+i\pi\sum_p\tilde\epsilon_p\overline n_p}. \nonumber \\ 
\end{eqnarray}
In deriving this, use has been made of the relation 
\begin{eqnarray}
  \label{eq:yyB}
  \sum_{a<b}B_{ab}y^ay^b&=&2\pi\sum_p\nu_p\xi^p\Biggl(\sum_q\gamma^{pq}\nu_q\xi^q+\eta^p\Biggr). 
\end{eqnarray}
The orthogonality of $\psi^{r_1\cdots r_n}$ with different set of $\{r_p\}$ is 
obvious by the standard integral 
\begin{eqnarray}
  \label{eq:onropw}
  \int_0^1d^n\eta e^{2\pi i\sum_p\overline n_p\eta^p}
  e^{-2\pi i\sum_q\overline n'_q\eta^q}
  =\delta_{\overline n_1,\overline n'_1}\cdots\delta_{\overline n_n,\overline n'_n}.
\end{eqnarray}
The computation of the normalization constant is given in Appendix \ref{sec:NIWF}. 

The zero-mode wave function (\ref{eq:zmwfix}) is the generalization of 
(\ref{eq:zms}) to arbitrary even dimensions. It is interesting to see how 
the periodicity conditions  (\ref{eq:psiagpc}), or their generalization for 
an arbitrary integral vector $L=\sum_aL^ae_a$ 
\begin{eqnarray}
  \label{eq:gpcpsi}
  \psi(x+L)&=&e^{i\sum_{a<b}F_{ab}L^ax^b}\psi(x), 
\end{eqnarray}
are satisfied. These can be verified by 
applying (\ref{eq:psiagpc}) successively to the lhs of (\ref{eq:gpcpsi}). Since the $2n$ 
vectors $M_p=(M_p{}^a)$ and $N_p=(N_p{}^a)$ span the entire lattice ${\bf Z}^{2n}$, it 
suffices to show (\ref{eq:gpcpsi}) for $L=M_p, ~N_p$. In the form given by 
(\ref{eq:zmwfix}) it is completely straightforward to check these. For instance, under 
the translation $x\rightarrow x+M_p$  the overall exponential factor of $\psi(x)$ 
develops an extra phase factor as 
\begin{eqnarray}
  \label{eq:oapf}
  e^{\frac{i}{2}\sum_{a<b}(x^a+M_p{}^a)(x+M_p{}^b)}&=&
  e^{-i\pi \nu_p\eta^p+i\pi\tilde\epsilon_p\nu_p}e^{i\sum_{a<b}M_p{}^ax^bF_{ab}}
  e^{\frac{i}{2}\sum_{a<b}x^ax^bF_{ab}}. 
\end{eqnarray}
On the other hand $F_{r_1\cdots r_n}(x)$ is transformed into 
\begin{eqnarray}
  \label{eq:trFM}
  F_{s_1\cdots s_n}^{r_1\cdots r_n}(x+M_p)&=&e^{i\pi\nu_p\eta^p+i\pi\tilde\epsilon_p\nu_p}
  F_{s_1\cdots s_n}^{r_1\cdots r_n}(x). 
\end{eqnarray}
Since $\tilde\epsilon_p\nu_p$ is an integer, the extra phase factors in the rhs' 
of (\ref{eq:oapf}) and (\ref{eq:trFM})  cancel each other. Similar thing also 
happens for the translation $x\rightarrow x+N_p$. 

We finally note that the 
transformation properties of $F_{r_1\cdots r_n}(x)$ under (\ref{eq:xtrMN}) 
become symmetric as 
\begin{eqnarray}
  \label{eq:trFN}
  F_{s_1\cdots s_n}^{r_1\cdots r_n}(x+M_p)&=&e^{\frac{i}{2}\sum_{a<b}F_{ab}(M_p{}^ax^b
    -M_p{}^bx^a+M_p{}^aM_p{}^b)}
  F_{s_1\cdots s_n}^{r_1\cdots r_n}(x), \nonumber \\
  F_{s_1\cdots s_n}^{r_1\cdots r_n}(x+N_p)&=&e^{\frac{i}{2}\sum_{a<b}F_{ab}(N_p{}^ax^b
    -N_p{}^bx^a-N_p{}^aN_p{}^b)}
  F_{s_1\cdots s_n}^{r_1\cdots r_n}(x). 
\end{eqnarray}
In deriving these we have used (\ref{eq:x2xieta}), (\ref{eq:ep}) and (\ref{eq:et}).

\section{Summary}
\label{sec:sandd}

We have investigated the Dirac operator zero-modes on a torus for a 
gauge background with constant field strength. The gauge potential possesses 
a characteristic transformation properties under the translations corresponding 
to the periods of the torus. This gives rise to nontrivial periodicity conditions for 
the wave function.  In the coordinates where the field strength take a standard 
block-diagonal form the Dirac equation for the zero-modes can be reduced to 
first order differential equations for each spinor component. It give a kind of 
holomorphicity conditions and  allows us to express the wave function by a set 
of holomorphic functions of the torus coordinates. We have shown that it 
is always possible to find a set of complex torus coordinates in which 
the holomorphic part of the wave function becomes either periodic or 
anti-periodic under unit translations along the imaginary axes. 
The periodicity conditions determine up to an overall normalization the 
holomorphic part of the wave function. For gauge backgrounds with 
uniform field strength only left-handed or right-handed chiral zero-modes 
are realized in a manner consistent with the index theorem. 

Throughout the paper we have assumed $\det F_{ab}\ne0$. If $\det F_{ab}=0$,  
then the index of the Dirac operator vanishes. We expect two possibilities. 
One is the occurrence of an equal number of positive and negative chirality 
zero-modes. Another is the absence of chiral zero-modes. As noted in 
Sect. \ref{sec:det2}, we cannot remove some of the constant components 
of the gauge potential by shifting the coordinate origin if $\det F_{ab}=0$. 
In general the presence of a constant gauge potential affects the spectrum 
of the Dirac operator and eliminates the chiral zero-modes. A complete 
analysis of the eigenvalue problem of the Dirac operator is interesting. 
We will argue this in a separate publication. 

\vskip .3cm
This work is supported in part by the Grant-in Aid for Scientific Research 
form the Ministry of Education, Culture, Sports, Science and Technology 
(No. 13135203). 

\appendix
\section{General Translations}
\label{sec:gt}
\setcounter{equation}{0}

We first consider a translation $z\rightarrow z+n\Xi_a$ with an arbitrary 
integer $n$. By a successive application of (\ref{eq:pc}), we find 
\begin{eqnarray}
  \label{eq:pc0}
  f_{s_1\cdots s_n}(z+n\Xi_a)&=&e^{\sum_kns_k\zeta_a^{2k-1}B_k\bigl(z^k
    +\frac{1}{2}n\Xi_a^k\bigr)}f_{s_1\cdots s_n}(z).
\end{eqnarray}
We now consider the most general translation by 
\begin{eqnarray}
  \label{eq:mgtrxi}
  \Xi^k=\sum_aN^a\Xi^k_a.
\end{eqnarray}
By applying (\ref{eq:pc0}) to 
\begin{eqnarray}
  \label{eq:xitmp10}
  f_{s_1\cdots s_n}\Bigl(z+\sum_{a=1}^{b}N^a\Xi_a\Bigr)
=f_{s_1\cdots s_n}\Bigl(z+\sum_{a=1}^{b-1}N^a\Xi_a+N^b\Xi_b\Bigr),
\end{eqnarray}
we obtain a recursive relation 
\begin{eqnarray}
  \label{eq:xi12k}
  f_{s_1\cdots s_n}\Bigl(z+\sum_{a=1}^{b}N^a\Xi_a\Bigr)&=&
  e^{\sum_kN^bs_k\zeta_b^{2k-1}B_k\Bigl(z^k+\sum_{a=1}^{b-1}N^a\Xi_a^k
    +\frac{1}{2}N^b\Xi_b^k\Bigr)}
  f_{s_1\cdots s_n}(z+\sum_{a=1}^{b-1}N^a\Xi_a). \nonumber \\
\end{eqnarray}
This gives 
\begin{eqnarray}
  \label{eq:gtrX1N}
  f_{s_1\cdots s_n}(z+\Xi)&=&
  \prod_{b=1}^{2n}\Biggl(e^{\sum_kN^bs_k\zeta_b^{2k-1}B_k
    \Bigl(z^k+\sum_{a=1}^{b-1}N^a\Xi_a^k+\frac{1}{2}N^b\Xi_b^k\Bigr)}\Biggr)
  f_{s_1\cdots s_n}(z) \nonumber \\
  &=&e^{\sum_ks_kB_k\bigl(\zeta^{2k-1}z^k+
    \frac{1}{2}\sum_a(N^a)^2\zeta_a^{2k-1}\Xi_a^k
    +\sum_{a>b}N^aN^b\zeta_a^{2k-1}\Xi_b^k\bigr)}f_{s_1\cdots s_n}(z),  \nonumber \\
\end{eqnarray}
where use has been made of $\zeta^{2k-1}=\sum_aN^a\zeta_a^{2k-1}$. 
By noting the relation  
\begin{eqnarray}
  \label{eq:szxi}
  \sum_a(N^a)^2\zeta_a^{2k-1}\Xi_a^k
    +2\sum_{a>b}N^aN^b\zeta_a^{2k-1}\Xi_b^k
    =\zeta^{2k-1}\Xi^k-is_k\sum_{a<b}N^aN^b(\zeta_a^{2k-1}\zeta_b^{2k}
    -\zeta_a^{2k}\zeta_b^{2k-1})
    \nonumber \\
\end{eqnarray}
and using (\ref{eq:F2bzz}), we see that (\ref{eq:gtrX1N}) can be written in the form 
(\ref{eq:fXitr}).

\section{Proof of the Lemma}
\label{sec:fl}
\setcounter{equation}{0}

In this appendix we give a proof of the basic lemma mentioned in Sect. \ref{sec:gpspc}. 
It can be stated in the following form;

\medskip\noindent
{\bf Lemma}: Any nondegenerate 
antisymmetric matrix $\phi$ with $\phi_{ab}=-\phi_{ba}\in{\bf Z}$ 
can be converted to a block-diagonal form 
\begin{eqnarray}
  \label{eq:frob}
  L_a{}^cL_b{}^d\phi_{cd}=\nu_{ab},\qquad\hbox{with} \quad
  \nu=\left(
  \begin{matrix}
    0&-\nu_1&&&& \\
    \nu_1&0&&&& \\
    &&0&-\nu_2&&& \\
    &&\nu_2&0&&& \\
    &&&&\ddots&&&\\
    &&&&&0&-\nu_n\\
    &&&&&\nu_n&0
  \end{matrix}\right),
\end{eqnarray}
where $L$ is a $2n\times 2n$ matrix with $L_a{}^b\in{\bf Z}$ and $|\det L|=1$. 
The $\nu_1,\cdots,\nu_n$ are a sequence of positive integers with 
$\nu_{p+1}/\nu_p\in{\bf Z}$. 

\medskip\noindent
To prove this let us consider a space ${\bf Z}^{2n}$ spanned by the vectors 
\begin{eqnarray}
  \label{eq:xieqxieea}
  \xi=\sum_a\xi^ae_a, \qquad (\xi^a\in{\bf Z}). 
\end{eqnarray}
Then $\phi$ naturally defines an integer-valued skew symmetric bilinear form 
\begin{eqnarray}
  \label{eq:phixieta}
  \phi(\xi,\eta)=\sum_{a,b}\phi_{ab}\xi^a\eta^b. \qquad 
  (\xi,~\eta\in{\bf Z}^{2n})
\end{eqnarray}
Let $M_1$ and $N_1$ be such a pair of vectors 
that $\nu_1=\phi(M_1,N_1)$ becomes the smallest positive integer. 
Then we see that $\phi(M_1,\xi)$ and $\phi(N_1,\xi)$ for any integral 
vector $\xi$ must be divided by $\nu_1$.  This can be shown as follows:
Let $\xi$ be such an integral vector with $\phi(M_1,\xi)/\nu_1\notin{\bf Z}$. 
Then we have  $\phi(M_1,kN_1+\xi)=k\nu_1+\phi(M_1,\xi)$ for any integer 
$k$. Since $|\phi(M_1,\xi)|>\nu_1$ by assumption, we can always find 
a suitable $k$ such that $0<\phi(M_1,kN_1+\xi)<\nu_1$ is satisfied. This 
contradicts the assumption that $\nu_1$ is the smallest positive 
integer taken by $\phi$. Hence $\phi(M_1,\xi)$ must be divided by 
$\nu_1$. This also applies to $\phi(N_1,\xi)$. 
Using these properties, we can decompose any integral vector $\xi$ 
in the form 
\begin{eqnarray}
  \label{eq:decxi}
  \xi=\xi'+\frac{\phi(\xi,N_1)}{\nu_1}M_1+\frac{\phi(M_1,\xi)}{\nu_1}N_1, 
\end{eqnarray}
where $\xi'$ satisfies 
\begin{eqnarray}
  \label{eq:soivxp}
  \phi(M_1,\xi')=\phi(\xi',N_1)=0.
\end{eqnarray}

We next consider the space of integral vectors $\xi'$ satisfying 
(\ref{eq:soivxp}) and evaluate the bilinear form $\phi$. 
Let $M_2$ and $N_2$ be  such a pair of vectors that $\nu_2=\phi(M_2,N_2)$ 
becomes the smallest positive integer  on this restricted space. By 
assumption we have $\nu_1\leq \nu_2$. To see that $\nu_2/\nu_1\in{\bf Z}$ 
we consider 
\begin{eqnarray}
  \label{eq:phiMN}
  \phi(kM_1+M_2,N_1+N_2)=k\nu_1+\nu_2,
\end{eqnarray}
where $k$ may be an arbitrary integer. Since $k\nu_1+\nu_2$ cannot be a 
positive integer smaller than $\nu_1$ for any integer $k$, we see that  
$\nu_1$ must divide $\nu_2$. We can also show that $\phi(M_2,\xi)$ and 
$\phi(N_2,\xi)$ for any vector $\xi\in{\bf Z}^{2n}$ can be divided by $\nu_2$. 

This procedure can be continued until we arrive at $M_n$ and $N_n$ with 
$\nu_n=F(M_n,N_n)$. The matrix $L_a{}^b$ is then given by 
\begin{eqnarray}
  \label{eq:L2MN}
  L_{2p-1}{}^a=N_p^a, \quad L_{2p}{}^a=M_p^a. \qquad (p=1,\cdots,n)
\end{eqnarray}
In matrix notation this can be written as
\begin{eqnarray}
  \label{eq:LeqN1M1}
  L=(N_1,M_1,\cdots,N_n,M_n).
\end{eqnarray}

To show $|\det L|=1$ we note that any vector $\xi\in{\bf Z}^{2n}$ can be 
expanded uniquely as
\begin{eqnarray}
  \label{eq:xieqsum}
  \xi=\sum_{p}(a^pN_p+b^pM_p) \quad
  \hbox{with}\quad a_p=\frac{1}{\nu_p}\phi(M_p,\xi), 
  \quad b_p=\frac{1}{\nu_p}\phi(\xi,N_p),
\end{eqnarray}
where the coefficients $a_p$ and $b_p$ are all integers. 
In particular we consider expansion of unit vectors $e_a$ with 
$(e_a)^b=\delta_a^b$ as 
\begin{eqnarray}
  \label{eq:ea}
  e_a=\sum_p(A_{ap}N_p+B_{ap}M_p),
\end{eqnarray}
where $A_{ap}$ and $B_{ap}$ are all integers. In terms of 
components we have 
\begin{eqnarray}
  \label{eq:sumAap}
  \sum_p(A_{ap}N_p{}^b+B_{ap}M_p{}^b)=\delta_a^b.
\end{eqnarray}
These can be expressed in a matrix form as
\begin{eqnarray}
  \label{eq:ABNM}
  \left(
    \begin{matrix}
      A_{11}&\cdots&A_{1n}&B_{11}&\cdots&B_{1n} \\
      \vdots&\ddots&\vdots&\vdots&\ddots&\vdots\\
      A_{2n1}&\cdots&A_{2nn}&B_{2n1}&\cdots&B_{2nn}
    \end{matrix}\right)\left(
  \begin{matrix}
    N_1{}^1 &\cdots&N_1{}^{2n} \\
    \vdots &\cdots&\vdots \\
    N_n{}^1&\cdots&N_n{}^{2n} \\
    M_1{}^1 &\cdots&M_1{}^{2n} \\
    \vdots &\cdots&\vdots \\
    M_n{}^1&\cdots&M_n{}^{2n}
  \end{matrix}\right)=1.
\end{eqnarray}
Taking the determinant of both sides of this expression, we 
find that 
\begin{eqnarray}
  \label{eq:detL}
  |\det L|=|\det(N_1,\cdots,N_n,M_1,\cdots,M_n)|=1. 
\end{eqnarray}
This completes the proof of the lemma.

From (\ref{eq:frob}) and (\ref{eq:detL}) we obtain $\det\nu=\det\phi$. 
This immediately gives 
\begin{eqnarray}
  \label{eq:nunu}
  \nu_1\cdots\nu_n=\frac{1}{n!}|\epsilon_{a_1b_1\cdots a_nb_n}
  \phi_{a_1b_1}\cdots \phi_{a_nb_n}|.
\end{eqnarray}

\section{Existence of a Standard Basis}
\label{sec:add2}
\setcounter{equation}{0}

Using the the lemma given in Appendix \ref{sec:fl}, we can always find $2n$ sets 
of integral vectors $N_p=(N_p{}^a)$ and $M_p=(M_p{}^a)$ 
satisfying 
\begin{eqnarray}
  \label{eq:Feqnu}
  && F(M_p,N_q)=\nu_p\delta_{pq}, \qquad
  F(M_p,M_q)=F(N_p,N_q)=0, \nonumber \\
  && |\det(N_1,M_1,\cdots,N_n,M_n)|=1, 
\end{eqnarray}
where $\nu_1,\cdots,\nu_n$ are positive integers 
with the property that $\nu_p$ divides $\nu_q$ 
for $p<q$. To make the paper self-contained we give a proof of 
the existence of a set of orthonormal basis vectors 
$(\zeta^1,\cdots,\zeta^{2n})$ that converts the field strengths into 
a standard form (\ref{eq:FeqRBR}) and satisfies the 
conditions (\ref{eq:nzetaeq0}).  We basically follow Ref. 
\cite{Sakamoto-Tanimura}.

We first note that an arbitrary vector $V\in{\bf R}^{2n}$ 
can be expanded in three ways as 
\begin{eqnarray}
  \label{eq:V}
  V=\sum_av^ae_a=\sum_a u^aL_a=\sum_aw^a\zeta^a, 
\end{eqnarray}
where the bases $e_a$, $L_a=(N_1,M_1,\cdots,N_n.M_n)$ 
and $\zeta^a$ in ${\bf R}^{2n}$ are defined by
\begin{eqnarray}
  \label{eq:eab}
  (e_a)^b=\delta_a^b, \qquad
  (L_a)^b=L_a{}^b, \qquad
  (\zeta^a)^b=\zeta_b^a. 
\end{eqnarray}
This gives 
\begin{eqnarray}
  \label{eq:va}
  w^c=\sum_{a,b}u^aL_a{}^b\zeta_b^c.
\end{eqnarray}
The transformation matrix 
\begin{eqnarray}
  \label{eq:LR}
  (LR)_a{}^b=\sum_cL_a{}^c\zeta_c^b
\end{eqnarray}
relates $B=(B_{ab})$ with $\nu=(\nu_{ab})$ as
\begin{eqnarray}
  \label{eq:nuLFLT}
  \nu=LFL^T=LRB(LR)^T. 
\end{eqnarray}
We then define $n$ dimensional vector subspaces spanned 
by $\{N_1,\cdots,N_n\}$ and $\{M_1,\cdots,M_n\}$ as
\begin{eqnarray}
  \label{eq:Mm}
  &&M^-=\{V\in R^{2n}|V=\sum_p v^pN_p,~v^p\in{\bf R}\}, \\
  &&M^+=\{V\in R^{2n}|V=\sum_p v^pM_p,~v^p\in{\bf R}\}.
\end{eqnarray}
The field strength 2-form (\ref{eq:fs2f}) 
is degenerated on $M^+$ and $M^-$ in the sense 
\begin{eqnarray}
  \label{eq:FFeq0}
  F(N_p,N_q)=F(M_p.M_q)=0. 
\end{eqnarray}
Furthermore, $M^+$ and $M^-$ are maximal degenerate 
subspaces of ${\bf R}^{2n}$. This implies $F(V,W)\ne0$ for 
some $V\in M^+$ and $W\notin M^+$. 

The ${\bf R}^{2n}$ can also be decomposed into $n$ 
two-dimensional orthogonal subspaces $W_1$, $\cdots$, 
$W_n$, where $W_k$ is spanned by the orthonormal 
basis vectors $\zeta^{2k-1}=(\zeta_a^{2k-1})$ and 
$\zeta^{2k}=(\zeta_a^{2k})$ satisfying 
\begin{eqnarray}
  \label{eq:FFz}
  \sum_{b,c}F_{ab}F_{bc}\zeta_c^{2k-1}=-B_k^2\zeta_a^{2k-1}, \qquad
  \zeta_a^{2k}=\frac{1}{B_k}\sum_bF_{ab}\zeta_b^{2k-1}.
\end{eqnarray}
We note that the vector subspaces $W_k^\pm$ of $W_k$ 
defined by 
\begin{eqnarray}
  \label{eq:Wkm}
  W_k^-=W_k\cap M^-, \qquad
  W_k^+=W_k\cap(M^-)^\perp
\end{eqnarray}
must be one-dimensional. This can be seen as follows;
If $\dim W_k^-=2$, we have $W_k^-=W_k\subseteq M^-$. 
Since $F$ is degenerated on $M^-$, we obtain 
$F(V,V')=0$ for arbitrary $V,V'\in W_k^-$. This 
contradicts the fact that the basis vectors $\zeta^{2k-1}$ and 
$\zeta^{2k}$ defined above satisfy 
\begin{eqnarray}
  \label{eq:2piFz}
  F(\zeta^{2k-1},\zeta^{2k})=-\frac{B_k}{2\pi}\ne0.
\end{eqnarray}
On the other hand if $\dim W_k^+=2$, we see that 
$W_k^+=W_k\subset (M^-)^{\perp}$ is orthogonal 
to $M^-$. Then we can expand $N_p$ as
\begin{eqnarray}
  \label{eq:Np}
  N_p=\sum_{l(\ne k)}(c_p^l\zeta^{2l-1}+d_p^l\zeta^{2l}). 
\end{eqnarray}
We now choose an arbitrary vector $V\in W_k^+$ and consider 
the $n+1$ dimensional subspace $\tilde M^-\supset M^-$ 
of ${\bf R}^{2n}$ spanned by $V$ and $\{N_1,\cdots,N_n\}$. 
Using (\ref{eq:Np}), we immediately find $F(V,N_p)=0$ for any $p$. 
This implies that $F$ is degenerated on $\tilde M^-$. 
This contradicts the fact that $M^-$ is a
maximal degenerate subspace. We thus conclude that 
$\dim W_k^\pm=1$. 

We can alway find an orthonormal basis $(\zeta^{2k-1},\zeta^{2k})$ 
in $W_k$ to satisfy
\begin{eqnarray}
  \label{eq:z2k-1}
  \zeta^{2k-1}\in W_k^-, \qquad \zeta^{2k}\in W_k^+. 
\end{eqnarray}
Since $W_k^-$ is orthogonal to $M^-$ we obtain 
\begin{eqnarray}
  \label{eq:Npaz}
  \sum_aN_p{}^a\zeta_a^{2k-1}=0. 
\end{eqnarray}
This completes the proof of the existence of $\zeta_a^b$ satisfying 
(\ref{eq:nzetaeq0}). 

\section{Computation of the Normalization Constant}
\label{sec:NIWF}
\setcounter{equation}{0}

In this appendix we compute the normalization constant of the 
wave function (\ref{eq:zmwfix}). Due to the orthogonality 
relations (\ref{eq:onropw}) we have only to compute the 
integral
\begin{eqnarray}
  \label{eq:normc}
  &&\int_Dd^{2n}x\Biggl|\sum_{\overline n_1,\cdots,\overline n_n}
  e^{-\pi\sum_{p,q}(g^{pq}+i\gamma^{pq})(\nu_p\xi^p-\overline n_p)(\nu_q\xi^q-\overline n_q)
  -i\pi\sum_p(\nu_p\xi^p-\overline n_p)\eta^p
  +i\pi\sum_p\overline n_p\eta^p+i\pi\sum_p\tilde\epsilon_p\overline n_p}\Biggr|^2 \nonumber\\
  &&\hskip 0.6cm =\int_0^1d^n\xi d^n\eta\sum_{\overline n_1,\cdots,\overline n_n}
  \sum_{\overline n'_1,\cdots,\overline n'_n}
  e^{-\pi\sum_{p,q}(g^{pq}+i\gamma^{pq})(\nu_p\xi^p-\overline n_p)(\nu_q\xi^q-\overline n_q)
  -i\pi\sum_p(\nu_p\xi^p-\overline n_p)\eta^p
  +i\pi\sum_p\overline n_p\eta^p+i\pi\sum_p\tilde\epsilon_p\overline n_p}\nonumber \\
  &&\hskip 5cm\times e^{-\pi\sum_{p,q}(g^{pq}-i\gamma^{pq})(\nu_p\xi^p-\overline n'_p)(\nu_q\xi^q-\overline n'_q)
  +i\pi\sum_p(\nu_p\xi^p-\overline n'_p)\eta^p
  -i\pi\sum_p\overline n'_p\eta^p-i\pi\sum_p\tilde\epsilon_p\overline n'_p} \nonumber\\
  &&\hskip 0.6cm =\int_0^1d^n\xi\sum_{\overline n_1,\cdots,\overline n_n}
  \sum_{\overline n'_1,\cdots,\overline n'_n}
  e^{-\pi\sum_{p,q}(g^{pq}+i\gamma^{pq})(\nu_p\xi^p-\overline n_p)(\nu_q\xi^q-\overline n_q)
  +i\pi\sum_p\tilde\epsilon_p\overline n_p} \nonumber\\
  &&\hskip 3cm\times e^{-\pi\sum_{p,q}(g^{pq}-i\gamma^{pq})(\nu_p\xi^p-\overline n'_p)(\nu_q\xi^q-\overline n'_q)
  -i\pi\sum_p\tilde\epsilon_p\overline n'_p}
  \int_0^1 d^n\eta e^{2i\pi\sum_p(\overline n_p-\overline n'_p)\eta^p} \nonumber \\  
  &&\hskip 0.6cm =\int_0^1d^n\xi\sum_{\overline n_1,\cdots,\overline n_n}
  e^{-2\pi\sum_{p,q}g^{pq}(\nu_p\xi^p-\overline n_p)(\nu_q\xi^q-\overline n_q)} \nonumber\\
  &&\hskip 0.6cm =\int_0^1d^n\xi\sum_{k_1,\cdots,k_n=-\infty}^{+\infty}
  e^{-2\pi\sum_{p,q}g^{pq}(\nu_p\xi^p-\nu_pk_p-r_p+\frac{1}{2}\epsilon_p)
    (\nu_q\xi^q-\nu_qk_q-r_q+\frac{1}{2}\epsilon_q)} \nonumber\\
  &&\hskip 0.6cm =\sum_{k_1,\cdots,k_n=-\infty}^{+\infty}
  \int_{-k_1}^{-k_1+1}d\xi^1\cdots \int_{-k_n}^{-k_n+1}d\xi^n
  e^{-2\pi\sum_{p,q}g^{pq}(\nu_p\xi^p-r_p+\frac{1}{2}\epsilon_p)
    (\nu_q\xi^q-r_q+\frac{1}{2}\epsilon_q)} \nonumber\\  
    &&\hskip 0.6cm=
    \int_{-\infty}^{+\infty}d\xi^1\cdots\int_{-\infty}^{+\infty}d\xi^n
    e^{-2\pi\sum_{p,q}g^{pq}\nu_p\nu_q\xi^p\xi^q} \nonumber\\
    &&\hskip 0.6cm=\frac{\sqrt{\det g_{pq}}}{2^{\frac{n}{2}}\nu_1\cdots\nu_n}.
\end{eqnarray}
This justifies the overall normalization of (\ref{eq:zmwfix}). 
In two dimensions we have $g_{11}=\nu$. The normalization constant 
appearing in (\ref{eq:fs}) is consistent with (\ref{eq:normc}).


\newpage
\begin{thebibliography}{99}
\bibitem{Atiyah-Singer} M. F. Atiyah and I. M. Singer, Bull. Amer. Math. Soc. {\bf 69} (1963) 422; 
Ann. Math. {\bf 87} (1968) 484, 546.
\bibitem{Adler} S. L. Adler, Phys. Rev. {\bf 177} (1969) 2426.
\newline
J. Bell and R. Jackiw, Nuovo Cim. {\bf 60A} (1969) 47.
\bibitem{Nielsen-Ninomiya} H. B. Nielsen and M. Ninomiya, Phys. Lett. {\bf 130B} (1983) 389.
\newline 
A. Manohar, Phys. Lett. {\bf 153B} (1985) 153.
\newline
T. Fujiwara and Y. Ohnuki, Prog. Theor. Phys. {\bf 76} (1986) 1182; {\bf 77} (1987) 1463.
\bibitem{Sakamoto-Tanimura} M. Sakamoto and S. Tanimura, J. Math. Phys. {\bf 44} (2003)  5042 
[{\tt hep-th/0306006}] .
\bibitem{Brown} E. Brown, Phys. Rev. {\bf 133} (1964) A1038.
\bibitem{Zak} J. Zak, Phys. Rev. {\bf 134} (1964) A1602, A1607. 
\bibitem{Tanimura} S. Tanimura, J. Math. Phys. {\bf 43} (2002) 5926 [{\tt hep-th/0205053}].
\bibitem{Igusa} J. Igusa, ``Theta Functions,'' Lemma 5, pp. 71-72, Springer-Verlag 
(Berlin) (1972).  


\end{thebibliography}
\end{document}